\documentclass[twocolumn,showpacs,preprintnumbers,amsmath,amssymb]{revtex4}
\usepackage{amssymb}
\usepackage{amsbsy}
\usepackage{epsf}
\usepackage[cp866nav]{inputenc}
\usepackage{revsymb}
\usepackage[T2A,OT1]{fontenc}
\usepackage{graphicx}

\newcommand{\be}{\begin{equation}}
\newcommand{\ee}{\end{equation}}
\newcommand{\bea}{\begin{eqnarray}}
\newcommand{\eea}{\end{eqnarray}}

\begin{document}

\title
{A temperature behavior of the frustrated translational mode of
adsorbate and the nature of the ``adsorbate--substrate''
interaction}

\author{V.V. Ignatyuk}
\affiliation{ Institute for Condensed Matter Physics, 1
Svientsitskii Street, 79011, Lviv, Ukraine}

\date{\today}

\begin{abstract}

{A temperature behavior of the frustrated translational mode
(T-mode) of a light particle, coupled by different regimes of
ohmicity to the surface, is studied within a formalism of the
generalized diffusion coefficients. The memory effects of the
adsorbate motion are considered to be the main reason of the
T-mode origin. Numerical calculations yield a thermally induced
shift and broadening of the T-mode, which is found to be linear in
temperature for Ohmic and super-Ohmic systems and nonlinear for
strongly sub-Ohmic ones. We obtain analytical expressions for the
T-mode shift and width at weak coupling for the systems with
integer ``ohmicity'' indexes $n=0\div 2$ in zero temperature and
high temperature limits. We provide an explanation of the
experimentally observed {\it blue}- or {\it red}-shifts of the
T-mode on the basis of a comparative analysis of two typical times
of the system evolution: a time of decay of the
``velocity--velocity'' autocorrelation function, and a correlation
time of the thermal bath random forces. A relation of the T-mode
to the multiple jumps of the adsorbate is discussed, and
generalization of conditions of the multiple hopping to the case
of quantum surface diffusion is performed.}

\end{abstract}

\maketitle

 \setcounter{equation}{0}

\section{Introduction}

Among a number of problems dealt with the surface diffusion, which
occurs either due to thermally activated \cite{PhysA195,Ferrando}
or tunnelling mechanisms \cite{Ferrando,single}, one can single
out the investigation of the frustrated translational mode
(T-mode) of the adsorbate \cite{T-mode-1,T-mode-2}. The T-mode is
known to appear when the adparticle moves between two stable
positions at the surface within one adsorption site (for instance,
between the atop position and the saddle point \cite{single}) due
to interaction with phonons or electronic subsystem of the
substrate. This motion of the adsorbate is being detected during
inelastic helium atom scattering (IHAS) \cite{IHAS-1,IHAS-2} as an
additional peak in the dynamic structure factor
\cite{T-mode-2,Sfactor-1,Sfactor-2}. The vibrational nature of the
T-mode allows to refer this excitation to the external (low
frequency) modes of the adsorbate \cite{exMode-1,exMode-2}, which
are more or less temperature dependent and dispersionless with
homogeneous broadening at low coverages \cite{T-mode-2}.

In the last years the T-mode became a subject of the intensive
studies of both experimentalists and theorists. For instance, in
the recent scanning tunnelling spectroscopy experiments
\cite{Komeda} the excitation of the T-mode in polyatomic molecules
CO/Pd(110) has been explained in terms of the energy transfer from
the C-O stretch mode to the frustrated translational one due to
the anharmonic coupling between normal modes. A certain intrigue
is introduced by the fact that the T-mode is observed even in such
``classical'' systems as CO/Cu(001) and CO/Pt(111)
\cite{Csystems}, though the underlying mechanism of the T-mode
formation is purely quantum mechanical: the transitions between
ground and excited states within a well. This stimulated a
creation of the models based on classical \cite{Csystems} or
quantum \cite{Micha19} Langevin equations, the transition state
theory (TST) \cite{Kramers}, where anharmonicity of the  lattice
potential is taken into account, leading to the temperature
induced shift and broadening of the T-mode \cite{T-mode-1}.

At the same time, a theoretical approach for the T-mode
description has been proposed \cite{T-mode-2} from purely quantum
mechanical concepts. This method allows to extrapolate the results
obtained at $T\ne 0$ to the zero temperature limit, yielding
information about the ``adsorbate-substrate'' interaction strength
and vibrational frequency from the IHAS data. Much effort has been
put \cite{T-mode-2} into description of the T-mode temperature
dependence. It is known that unlike diffusion the vibrational
damping is not an activated process \cite{T-mode-2,JCP2}.
Nevertheless, the temperature induced T-peak shifts towards lower
frequencies, and almost linear broadening is observed
\cite{Csystems} for the Na/Cu(001) system in a region between 0 K
and 200 K, while for the CO/Cu(001) and CO/Pt(111) systems the
T-mode is {\it blue}-shifted with temperature. It is believed that
at low temperatures the shape of the T-mode is dominated by a
frictional damping, characterized by the nonadiabatic coupling to
the substrate excitations \cite{T-mode-2}. At high temperatures
the anharmonicity of the static lattice potential plays a dominant
role \cite{T-mode-1}. Summarizing, it could be stated that up to
now there is no unique viewpoint which factors (phonon and/or e-h
excitations, anharmonic contributions, or memory effects) are
predominant in the T-mode development at different temperatures,
couplings and interaction regimes.

Last but not least, one can relate the existence of the T-mode to
multiple jumps of the adsorbate. The onset of multi-\-hops was
associated with an inelastic peak of the dynamic structure factor
at frequency $\omega_{osc}$ of the adparticle oscillations at the
well bottom \cite{PhysA195,SurScience311}. If the T-mode is well
resolved on the dynamic structure factor and not overlapped by the
central quasi-elastic peak, it could be said that thermalization
of the adsorbate velocity has not taken place yet, and multiple
jumps contribute significantly to the diffusion coefficient.

In the present paper we study the conditions of the T-mode
formation, its temperature dependence, and its connection with the
nature of the ``adsorbate--substrate'' interaction. We explore a
two-level model of the light particle coupled non-adiabatically to
the surface phonons \cite{JCP1,PRE2009,PRE2011}. The other kinds
of ``adsorbate--substrate'' interaction, like electronic friction
or nonlinear phonon coupling, will be also considered. The
underbarrier hopping to the nearest adsorption sites is postulated
to be a ``driving force'' of the surface motion of the adsorbate.

We consider non-Markovian effects of the adsorbate motion, dealt
with retarded relaxation of the lattice excitations, to be the
main reason of the T-mode origin. Since this one-particle
excitation is very sensitive to the form of the kinetic kernels,
rigorous numerical evaluation of the T-mode temperature behavior
is carried out without any approximation (e.g., like
Wigner-Weisskopff one \cite{T-mode-2,Micha19}). On the other hand,
both zero temperature and semiclassical limits for the kinetic
kernels are considered in our paper that allows us to obtain the
analytical expressions for the T-mode characteristics at low
``adsorbate--substrate'' coupling.

We perform our analysis in terms of the generalized diffusion
coefficients (which are related to the ``velocity--velocity''
autocorrelation functions) rather than dynamic structure factor.
The latest approach is often complicated due to the overlap of
quasi-elastic and inelastic peaks of the dynamic structure factor
\cite{PhysA195}, while the method of generalized diffusion
coefficients allows one to study the T-mode in more detail, being
focused on the adsorbate motion at the well bottom. We propose a
physical interpretation of the observed temperature induced shift
of the T-mode on the basis of a comparative analysis of two
typical timescales of the system dynamics: a time of decay
$\tau_v$ of the generalized diffusion coefficients and a
correlation time $\tau_{cor}$ of the thermal bath random forces.
The obtained results reveal a close connection between the
direction of the T-mode shift and the nature of the
``adsorbate--substrate'' interaction.

Our paper is organized in the following way. In Sec. II we present
the generalized diffusion coefficient of a light particle derived
by the method of quantum kinetic equations \cite{PRE2009,PRE2011}
on the basis of two-level dissipative model of adparticle coupled
to substrate phonons. In Sec. III we consider different
correlation functions of the thermal bath random forces, which
depend on the nature of the ``adsorbate--substrate'' interaction.
In the next Section we attribute the T-mode of the adsorbate to
the coherent part of the generalized diffusion coefficient and
calculate its temperature dependence numerically at different
values of ohmicity index $n$. A comparison of the obtained results
with the experimental data and the results of similar theoretical
approaches is provided in this Section. In Sec.~V we calculate the
T-mode frequency and width analytically in zero and high
temperature limits for strongly sub-Ohmic (ohmicity index $n=0$),
Ohmic, and the super-Ohmic systems with $n=2$. In Sec. VI,
according to Refs. \cite{PhysA195,SurScience311}, we relate the
T-mode to the onset of multiple jumps of the adsorbate and
generalize the conditions of the multiple hopping to the case of
quantum surface diffusion. In the last Section we discuss briefly
the obtained results and draw final conclusions.

\section{Generalized quantum surface diffusion coefficients}

\setcounter{equation}{0} We consider a light particle, which
performs underbarrier hopping to the nearest adsorption site,
oscillates between ground and excited states within the potential
well, and interacts with the lattice vibrations. Hamiltonian of
the system is chosen as follows ~\cite{JCP1,JCP2} \bea\label{H}
H=H_A+H_{int}+H_B, \eea where the adsorbate is described by the
two-band constituent \bea\label{H_A} H_A\!\!=\!\!\! \sum_{\langle
ss'\rangle}(-t_ 0 a^{\dagger}_{s 0}a_{s' 0}+t_1 a^{\dagger}_{s
1}a_{s' 1})\! +\!\! \sum_s\!\frac{\hbar\Omega}{2}(n_{s 1}-n_{s
0}). \eea Here $s$ denotes the site of the lattice; 0 and 1 are
the ground and excited states within a given well, and $\langle
ss'\rangle$ denotes a sum over the nearest-neighbor sites. The
quantum states within a well are referred to as ``vibrational''
ones with the vibrational frequency $\Omega$. $a^{\dagger}_{s i}$
($a_{s i}$) creates (destroys) a particle on the site $s$ in the
vibrational state $i$; $n_{s i}=a^{\dagger}_{s i}a_{s i}$ is the
number operator for this state, and $n_{s}=n_{s 0}+n_{s 1}$.
Hereafter we will deal with a single adparticle only, hence the
adparticle statistics becomes irrelevant. $t_0$ and $t_1$ are the
nearest-neighbor tunnel\-ling amplitudes in the ground and the
first excited states, respectively, and we expect that $t_1\gg
t_0$.

The coupling to phonons is considered to be local within each
well. Phonons may couple both to the adsorbate density operators
and to the vibrations within a quantum well. The interaction
Hamiltonian is \cite{JCP1}
\bea\nonumber\label{H_int}&&H_{int}=\sum_{s}\!\left\{\! n_
s\sum_q\gamma_{s q}(b_q+b^{\dagger}_q)+(a^{\dagger}_{s 0}a_{s 1}+
a^{\dagger}_{s 1} a_{s 0})\right.
\\
&&\left.\times\sum_q \chi_{s q}(b_q+b^{\dagger}_q)\!\right\}, \eea
where $b^{\dagger}_q$ ($b_q$) creates (destroys) a phonon with a
normal mode frequency $\omega_q$. The strengths $\gamma_{s q}$
($\chi_{s q}$) describe coupling of phonons to the density
(oscillation) modes of the adsorbate. The bandwidths $t_0$, $t_1$
and vibrational frequency $\Omega$ can be evaluated in the
framework of the eigenvector-eigenvalue problem for a periodic
potential, felt by an adsorbate due to the static lattice. The
coupling strengths are expressed via the mean values
$\Gamma=\langle s,i|V_{int}^s|s,j\rangle$ of the lattice
distortion potential $V_{int}^s$ over the localized Wannier states
$|s,j\rangle$ times the phase factor depending on the site number
$s$ and wave-vector $q$ \cite{JCP1}. Likewise, we suppose $\Gamma$
to be the same for different quantum states $ \{i,j\}=\{0,1\}$ and
use the dimensionless coupling parameter \bea\label{Gamma}
G=\frac{\Gamma^2}{M\omega^3_{max}} \eea to characterize the
``adsorbate--substrate'' interaction. Here $\omega_{max}$ stands
for the Debye frequency, and $M$ denotes the mass of the substrate
atom.

The last term in Eq.~(\ref{H}) \bea\label{H_B}
H_B=\sum_q\hbar\omega_q b^{\dagger}_q b_q \eea corresponds to the
phonon bath; longitudinal acoustic phonons only are taken into
account in this model.

Since the tunnelling amplitudes are always much smaller than the
coupling strengths, the interaction part of Hamiltonian
(\ref{H_int}) cannot be considered as a perturbation and has to be
taken into account exactly. The most refined method consists in
performing unitary transformations in order to exclude the linear
over the interaction terms from the system Hamiltonian. In the
unitary transformed Hamiltonian the tunnelling processes can be
considered as those with emmision/adsorption of the vir\-tual
phonons, and the same is true for the vibrational processes.

It is also useful to pass to the hybrid set of states for each
site: \bea\label{a_LR} a_{s {L\atop R}}\equiv \frac{1}{\sqrt
2}(a_{s 0}\pm a_{s 1}),\eea and similarly for the creation
operators. The designation $L$ or $R$ means that a single
adparticle is now localized on the left or right side of the given
well. We will refer to the vibrational transitions with $i\ne j$,
$\{i,j\}=\{L,R\}$, as the end-changing processes, supplying them
afterwards by the subscript (c), and transitions with $i=j$ will
be termed as the end-preserving ones with the corresponding
subscript (p).

Using the method of the reduced density matrix \cite{MorozovBook}
it is possible \cite{JCP2,PRE2009,PRE2011} to obtain the chain of
quantum kinetic equations for diagonal
$f_{s,s}(t)=\sum_{i=L,R}\langle a^{\dagger}_{s i} a_{s
i}\rangle^t_S$ and off-diagonal $f_{s,s'}(t)=\sum_{i=L,R}\langle
a^{\dagger}_{s' i} a_{s i}\rangle^t_S$ one-particle
non-equilibrium distribution functions, where the averaging is
taken with the statistical operator $\rho_S(t)$ of the adsorbate.
These integro-differential equations are linear in $f_{s,s}(t)$,
$f_{s,s'}(t)$ but are non-local in time; hence it is useful to
perform the Laplace transformation $\tilde
f(z)=\int_0^{\infty}\exp(-z t)f(t) dt$. Solving the equations for
the off-diagonal distribution functions and inserting the obtained
results into the equation for the diagonal ones, one can obtain
\cite{PRE2009,PRE2011} an expression for the generalized
(frequency dependent) diffusion coefficient: \bea\label{Dgen}
&&\!\!\!\!\!\!\!\tilde D(z)=\tilde D_{coh}(z)+D_{in}(z)\nonumber
\\
&&\!\!\!\!\!\!\!\!\!\!= \frac{a^2}{4}\!\left[\frac{2
t_{inter}^2/\hbar^2}{z+\tilde{\gamma}_{inter}(z)+\tilde{\gamma}_{intra}(z)+\tilde{\gamma}_{LL}^+(z)}
+\tilde{\gamma}_{inter}(z)\right]. \eea The first term $\tilde
D_{coh}(z)$ in (\ref{Dgen}) describes a coherent contribution to
the generalized diffusion coefficient. It can be interpreted
\cite{JCP2,PRE2009} in terms of a simple model of band-type motion
limited by scattering from the lattice (with interatomic spacing
$a$) at temperatures large relative to the bandwidth $t_{inter}$,
which is narrowed due to polaronic effect. The kinetic kernel
\bea\label{GamInter}
\tilde{\gamma}_{inter}(z)=4\tilde{\gamma}_{LL}(z)+2\tilde{\gamma}_{LR}(z)+2\tilde{\gamma}_{RL}(z)\eea
corresponds to the dissipative intersite motion of the adsorbate
and describes processes, when the adparticle performs random
site-to-site hopping (with or without the change of its quantum
state) owing to the interaction with the bath. The kinetic kernel
$\tilde{\gamma}_{intra}(z)$ in Eq.~(\ref{Dgen}) describes a
dissipative intrasite dynamics, when the adsorbate during its
scattering from the lattice gets enough energy from the bath to be
excited from the ground state to the upper level within the same
adsorption site (the opposite process of particle de-excitation
with a phonon emission is also taken into consideration).

The second term $\tilde{D}_{in}(z)$ in Eq.~(\ref{Dgen}) is an
incoherent contribution to the generalized diffusion coefficient.
This is the result expected from the random walk model for
diffusion with site-to-site hopping rate
$\tilde{\gamma}_{inter}(z)$, describ\-ing processes of the surface
phonon creation/annihilation, when the particle performs a
transition from one Wannier state to another.

The rates $\tilde{\gamma}_{intra}(z)$, $\tilde{\gamma}_{inter}(z)$
can be obtained from the Laplace transformation of the kinetic
kernels \be\label{Gamx} \gamma_x(\tau)=\omega_{max}\lambda_x^2
\mbox{Re}\{\exp[
-(\varphi_x(0)-\varphi_x(\tau))]-\exp[-\varphi_x(0)]\}, \ee
\be\label{GamLLPlus} \gamma_{LL}^+(\tau)\!=\omega_{max}t_{1}^2
\mbox{Re}\{\exp[
-(\varphi_{LL}(0)+\varphi_{LL}(\tau))]-\exp[-\varphi_{LL}(0)]\},\ee
where \bea\label{phi}
\varphi_x(\tau)\!=\!\!\!\int\limits_{0}^1\!\!\frac{J_x(\omega)}{\omega^2}\!\left[\coth\!\left(
\frac{\hbar\omega}{2 k_B T}\!\right)\!\cos(\omega
\tau)-i\sin(\omega \tau) \!\right],\eea and one-particle
parameters $\lambda_x=\{t_1,\Omega\}$ are related to the
corresponding end-changing/end-preserving processes (see also
Table in Ref.~\cite{PRE2009}). The exponential form of the rates
(\ref{Gamx})-(\ref{GamLLPlus}) is a result of averag\-ing of the
lattice time correlation functions \cite{JCP2,PRE2009} over the
bath variables that yields the temperature dependent factor in
Eq.~(\ref{phi}). Hereafter we use dimensionless frequencies in the
units of $\omega_{max}$ and temperatures in the units of
$\hbar\omega_{max}/k_B$.

The ``adsorbate--substrate'' interaction enters the kinetic rates
via spectral weight functions \cite{JCP1,JCP2,PRE2009,PRE2011}
\be\label{Jintra} J(\omega)=\sum\limits_q\chi^2_{s
q}\delta(\omega-\omega_q), \ee \vspace*{-5mm}
 \bea\label{Jc}\nonumber
J_{LR}(\omega)\!=\!\!\sum\limits_q\left[(\gamma_{s q}-\gamma_{s'
q})\!+\!(\chi_{s q}+\chi_{s' q})
\right]^2\!\delta(\omega-\omega_q),
\\
\\
\nonumber J_{RL}(\omega)\!=\!\!\sum\limits_q\left[(\gamma_{s
q}-\gamma_{s' q})\!-\!(\chi_{s q}+\chi_{s' q})
\right]^2\!\delta(\omega-\omega_q),
 \eea
\bea\label{Jp}\nonumber
J_{LL}(\omega)\!=\!\!\sum\limits_q\left[(\gamma_{s q}-\gamma_{s'
q})\!+\!(\chi_{s q}-\chi_{s' q})
\right]^2\!\delta(\omega-\omega_q),
\\
\\
\nonumber
J_{RR}(\omega)\nonumber\!=\!\!\sum\limits_q\left[(\gamma_{s
q}-\gamma_{s' q})\!-\!(\chi_{s q}-\chi_{s' q})
\right]^2\!\delta(\omega-\omega_q).
 \eea
The function (\ref{Jintra}) describes the intrasite dynamics; the
functions (\ref{Jc}) are related to the intersite end-changing
processes, while (\ref{Jp}) are dealt with the intersite
end-preserving processes. The spectral weight function
(\ref{Jintra}) can be considered as site-independent if the system
has a translational symmetry, whereas the functions
(\ref{Jc})-(\ref{Jp}) depend only on the distance $|s-s'|$ between
the nearest neighbor sites $s$ and $s'$ \cite{JCP1,PRE2011}.

The system dynamics is governed by the low frequency behavior of
the spectral weight functions (\ref{Jintra})-(\ref{Jp}). In
Refs.~\cite{JCP1,JCP2,PRE2009,PRE2011} they were chosen to be
scaled as \bea\label{JJc} J_c(\omega)\approx
\Theta(\omega-\omega_0)\Theta(\omega_{max}-\omega)\eta_c\omega^{D-2},
\eea for the end-changing, and \bea\label{JJp}
 J_p(\omega)\approx
\Theta(\omega-\omega_0)\Theta(\omega_{max}-\omega)\eta_c\omega^{D},\eea
for the end-preserving processes with $\eta_c=10 G$, $\eta_p=12.5
G$, and $\Theta(x)$ denoting the Heaviside step function.

In Eqs.~(\ref{JJc})-(\ref{JJp}) the power index $D$ equals the
system dimensionality. Thus, for a two-dimensional lattice the
end-changing spectral weight functions are sub-Ohmic, while the
end-preserving ones are super-Ohmic \cite{Leggett}. Such a
behavior is a consequence of the equality of the coupling
strengths $\Gamma$ for all the quantum states. Moreover, the
lattice is assumed to possess a nonzero lowest frequency
$\omega_0$, which is introduced to take into account the finite
size of the system. It not only removes the divergencies
\cite{Leggett}, when the sub-Ohmic spectral functions have the
power index $n = 0$, but also allows one to describe the
adsorbate-induced surface reconstruction \cite{JCP1}, when the
particles become self-trapped due to the overlap of lattice
distortions.

We would like to note that a sharp cut-off of the spectral weight
functions at the Debye frequency $\omega_{max}$ is not a unique
one, and plain exponential \cite{Leggett} or algebraic
\cite{Weiss} cut-offs are also used. However, the obtained results
are found to be rather insensitive to the cut-off form, being
governed above all by the low frequency behavior of $J(\omega)$.

In the next Section we will consider other values of the power
index $n$ in the expressions for spectral weight functions and
discuss their relation to the processes of different physical
nature. Here we would like to emphasize that there is a
one-\-to-\-one correspondence between the low frequency behavior
of $J(\omega)$ and the long-time relaxation of the kinetic kernels
(\ref{Gamx}). It was pointed out in Ref.~\cite{PRE2011} that at
large times we pass from a Gaussian form for $\gamma_x(\tau)$ (for
the sub-Ohmic processes with $n=0$) through exponential relaxation
of kinetic kernels (for the Ohmic processes) to the power law
behavior (in the super-Ohmic case with $n=2$).

To conclude this Section, let us note that at a reasonable
assumption of an infinitesimal value of the tunnel\-ling amplitude
in comparison with vibrational frequency, $t_{1}\ll\hbar\Omega$,
the expression (\ref{Dgen}) for the generalized diffusion
coefficient can be simplified and presented as follows
\bea\label{Dgen1} &&\tilde D(z)\approx \frac{a^2
t_1^2}{2\hbar^2}\left[\frac{\exp(-\varphi_p(0))}{z+\Omega^2\tilde{\gamma}_{c}(z)}
+2\left(\tilde{\gamma}_{c}(z)+\tilde{\gamma}_{p}(z)\right)\right],
\eea where $\tilde{\gamma}_c(z)$ and $\tilde{\gamma}_p(z)$ mean
the Laplace transforms of the end-changing (evaluated with
spectral weight functions (\ref{JJc})) and end-preserving
(evaluated with spectral weight functions (\ref{JJp})) kernels
(\ref{Gamx}) times factor $1/\omega_{max} \lambda_x^2$. A generic
form for the coherent contribution to $\tilde D(z)$ has a
structure similar to Eq.~(\ref{Dgen1}) even if the other kinds of
interaction (like electronic friction and/or anharmonic coupling
to phonon subsystem) are taken into account. In the very general
case, a denominator in the expression for $\tilde D_{coh}(z)$ will
consist of several terms of different physical origin, which
describe the vibrational transitions of the adsorbate within the
well. In Sec.~IV we attribute the T-mode to the coherent part of
the generalized diffusion coefficient, and explore
Eq.~(\ref{Dgen1}) to study in detail its features at different
types of the ``adsorbate--substrate'' interaction.

\section{Correlation functions of the thermal bath random forces}

\setcounter{equation}{0} We can look at the problem of quantum
surface diffusion from another viewpoint. Making use of the
solution of the Heisenberg equations of motion for the external
degrees of freedom \cite{Chaos27} one can derive the so-called
generalized quantum Langevin equation \cite{T-mode-2,Chaos}
\bea\label{GQLE}
M\ddot{q}(t)+M\int\limits_{t_0}^t\gamma(t-\tau)\dot{q}(\tau)d\tau+\frac{d
V(q,t)}{dq}=F_r(t) \eea for the particle with the mass $M$ and a
generalized coordinate $q(t)$ moving within the potential
$V(q,t)$, where the damping kernel $\gamma(t)$ is expressed via
spectral weight function in the following way \bea\label{dkernel}
\gamma(t)=\int\limits_{0}^{\infty}\frac{J(\omega)}{\omega}\cos(\omega
t)d\omega.\eea The mean value of the operator-valued noise
$F_r(t)$ is zero \bea\label{moment1} \langle
F_r(t)\rangle_{bath}=0, \eea whereas the time correlation function
of the thermal bath random forces can be expressed as follows
\bea\label{moment2}\nonumber S_{FF}(t)=\frac{1}{2}\langle
F_r(t)F_r(0)+F_r(0)F_r(t)\rangle_{bath}\\
=\frac{M}{\pi}\int\limits_0^{\infty}\frac{J(\omega)}{\omega}\hbar\omega\coth\left(\frac{\hbar\omega}{2
k_B T}\right)\cos(\omega t)d\omega.\eea In the classical limit it
converts to the non-Markovian Einstein relation
\bea\label{moment2M} S_{FF}^{cl}(t)=M k_B T \gamma(t), \eea where
the average in Eqs.~(\ref{moment1})-(\ref{moment2}) is taken with
respect to a bath density matrix, which contains shifted
oscillators \cite{Chaos}.

The relation (\ref{moment2}) implies that modelling of quantum
dissipation in the system is possible in terms of macroscopic
quantities such as the friction kernel $\gamma(t)$ and temperature
$T$. A class of such systems is by no means restricted to the
``adparticle -- linear Debye phonons'' coup\-ling. One can give
examples of several types of the ``adsorbate-substrate'' coupling,
which are closely related to the low frequency behavior of the
spectral weight functions. They range from the systems with a
flicker noise \cite{SubOhmic,YK} with $J(\omega)\sim\omega^0$
(like the end-changing spectral functions (\ref{JJc})) through the
Ohmic coupling with $J(\omega)\sim\omega^1$, when one takes into
account the electronic friction and/or anharmonic terms in
``adsorbate-substrate'' interaction \cite{MorozovBook,Weiss}, to
the super-Ohmic systems with $J(\omega)\sim\omega^2$, when
coupling to surface Debye phonons is considered
\cite{JCP1,PRE2011} (see also the end-preserving spectral
functions (\ref{JJp})).

At any ohmicity it is useful to introduce the correlation time
$\tau_{cor}$ which defines the decay rate of correlation functions
(\ref{moment2})-(\ref{moment2M}) of the thermal bath random
forces. In other words, at times about $\tau_{cor}$ the excess
energy of the lattice relaxes. For instance, in the systems with
flicker noise, the corresponding time correlation functions
(\ref{moment2M}) have the residual correlations
$\tau_{cor}\sim\omega_0^{-1}$ in damping kernels,
\bea\label{Tcorr0}\gamma(t)\sim-\eta_c Ci(\omega_0 t), \quad
Ci(z)=\int\limits_z^{\infty} dt\frac{\cos t}{t}.\eea The sub-Ohmic
case (\ref{Tcorr0}) with the power index $n=0$ implies an
introduction of the lower limit $\omega_0$ of frequency,
$0<\omega_0\ll\omega_{max}$, which is related to the inverse time
of the experiment duration \cite{YK}, or dealt with the finiteness
of the system  size \cite{JCP1,PRE2009,PRE2011} (see also
Eqs.~(\ref{JJc})-(\ref{JJp}) and subsequent explanation).

The Ohmic systems have a white-noise-like behavior $\gamma(t)\sim
\eta_c\delta(t)$, and there are ultra-short range time
correlations in the super-Ohmic case. The super-Ohmic coupling
with the power indexes $n>2$ is beyond our consideration. The
cases of non-integer $n$, which are typical for fractal systems,
will be considered solely for the reason of proper convergence of
the numerical calculations. In particular, in the next Section we
consider the ohmicity indexes $n=\epsilon$, and $n=2-\epsilon$
with $\epsilon\ll 1$. An alternative way is to use the lower bound
cut-off frequency $\omega_0$, as it is done in Sec.~V, where the
analytical expressions for the T-mode features in the
quasi-classical limit are obtained.

 There is also another typical time scale of
the system dynamics -- the relaxation time $\tau_v$ of the
velocity autocorrelation functions. This time will be introduced
in the next Section, and along with $\tau_{cor}$ will be used for
explanation of the direction of the temperature induced shift of
the T-mode on the basis of a comparative analysis between
$\tau_{cor}$ and $\tau_v$.

\section{T-mode as a coherent part of the generalized diffusion coefficient}

\setcounter{equation}{0}
As has been already said, the T-mode of
the adsorbate is being detected in IHAS experiments as an
additional (inelastic) peak in the dynamic structure factor. Thus,
to predict theoretically the T-mode features one has to evaluate
the dynamic structure factor \bea\label{Sfactor}\nonumber S({\bf
k},\omega)=\frac{1}{2\pi}\int\limits_{-\infty}^{\infty}
\exp^{-i\omega t}I({\bf k},t)dt
\\
=\frac{1}{2\pi}\int\limits_{-\infty}^{\infty} e^{-i\omega t}
\langle e^{-i{\bf k}\cdot{\bf R}(0)} e^{i{\bf k}\cdot{\bf
R}(t)}\rangle dt, \eea which is nothing but the Fourier transform
of the intermediate scattering function $I({\bf k},t)$. In
Ref.~\cite{T-mode-2} the model of a damped quantum harmonic
oscillator bilinearly coupled to a bath of lattice oscillators has
been proposed. Non-Markovian equations of motion for the
creation/anihilation operators of the adsorbate have been derived,
which have a form of the generalized Langevin equation
(\ref{GQLE}). The dynamic structure factor of the adsorbate has
been found with taking into consideration the memory effects.
Temperature dependences of the frequency of the inelastic peak of
$S({\bf k},\omega)$ and its full width on half maximum (FWHM) have
been studied in detail.

Though an investigation of the T-mode features using the dynamic
structure factor is closely related to scattering experiments,
this approach can sometimes be complicated due to the overlap of
quasi-elastic and inelastic peaks. It happens, for instance
\cite{PhysA195}, in the quasi-continuous regime of adsorbate
motion, at small fric\-tion and high values of $k$, when the
quasi-elastic peak of $S({\bf k},\omega)$, dealt with diffusion of
the adsorbate, becomes low and broad and cannot be well separated
from the inelastic peak at the vibrational frequency $\Omega$. The
inelastic peak also becomes indistinguishable at strong coupling
due to the impossibility of completing coherent oscillations in
the wells.

In such a case, investigation of the T-mode features using the
Fourier transform $\tilde
C_{vv}(\omega)=1/2\pi\,\mbox{Re}\left[\int_{-\infty}^{\infty}e^{i\omega
t}C_{vv}(t) dt \right]$ of the velocity autocorrelation function
$C_{vv}(t)$ looks more promising, being focused on the adsorbate
motion at the bottom of the potential well. For instance, $\tilde
C_{vv}(\omega)$ reveals the adsorbate motion in the potential
wells also in the overdamped regime at high barriers, when, in
spite of large friction, the particle can perform small parts of
oscillations with unthermalized velocity \cite{PhysA195}. On the
other hand, $\tilde C_{vv}(\omega)$ can be easily obtained from
the dynamic structure factor by a simple relation
\cite{SurScience311} \bea\label{CtoS}\tilde
C_{vv}(\omega)=\omega^2\lim_{k\to 0}\frac{S({\bf k},\omega)}{{\bf
k^2}}.\eea

It has been pointed out in our previous papers
\cite{PRE2009,PRE2011} that the generalized diffusion coefficients
are directly related to the velocity autocorrelation function
$C_{vv}(t)$, determined at the adsorption site $s$. We will follow
this rule, having dealt a time-dependent diffusion coefficient
$D(t)$ with $C_{vv}(t)$. Such a definition of the gene\-ralized
diffusion coefficient differs from that used in Ref.~\cite{aging},
where the relation $\bar D(t)=\int_0^t C_{vv}(u)du$ has been
adopted. However, it is much more useful since it allows one to
associate the time of decay of $D(t)$ with the relaxation time
$\tau_v$ of the velocity autocorrelation function. On the other
hand, the zero frequency limit $D_{exp}=\pi\lim_{\omega\to
0}\tilde C_{vv}(\omega)=\int_0^{\infty} D(t)dt$ determines the
experimentally measured diffusion coefficients $D_{exp}$, whose
temperature dependence has been studied quite profoundly
\cite{Ferrando,single}.

In Ref.~\cite{PRE2011} it has been shown that there is a close
relation (at least, within the model proposed) between the
adparticle dynamics at intermediate times $\tau\ll\tau_v$ and
temperature dependence of the diffusion coefficients. Namely, as
the coupling strength increases, the adparticle motion (initially
oscillatory) becomes more and more smooth, indicating that the
temperature behavior of the diffusion coefficients $D_{exp}(T)$
should change from weakly dependent on $T$ to quite a sensitive
function of temperature. The oscillatory behavior of the
generalized diffusion coefficient has been related to the T-mode
onset. In the frequency representation, oscillations of
$C_{vv}(t)$ yield the side wings of $\mbox{Re}[\tilde
C_{vv}(\omega)]$, localized in the vicinity of the vibrational
frequency $\Omega$. The T-mode frequency was found to increase
with temperature (see Fig.~3 in Ref.~\cite{PRE2011}). The
corresponding side peak of $\mbox{Re}[\tilde C_{vv}(\omega)]$
should be {\it blue}-shifted too.

\begin{figure}[htb]
\centerline{\includegraphics[height=0.27\textheight]{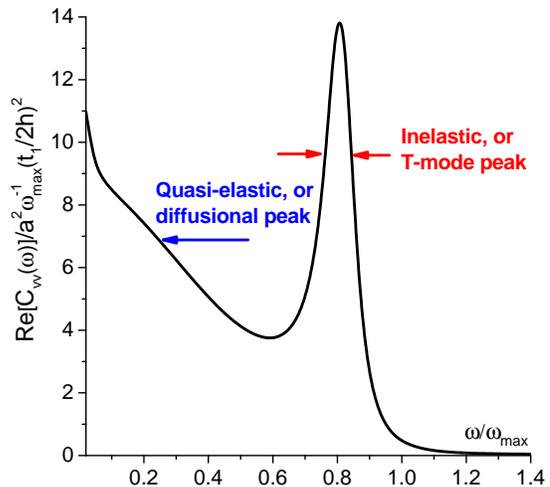}}
\caption{Frequency dependence of the Fourier
transform of velocity autocorrelation function at Debye
temperature $T_D= \hbar\omega_{max}/k_B$, ohmicity index $n=0$,
and parameters $t_{1}/\hbar\omega_{max}=10^{-5}$,
$\omega_0/\omega_{max}=10^{-4}$, $\eta_c= 10^{-2}$,
$\Omega/\omega_{max}=1/\sqrt{2}$.} \label{Tmode(omega)}
\end{figure}

Thus, there are two peaks of different physical origin in the
Fourier transform $\mbox{Re}[\tilde C_{vv}(\omega)]$ of the
velocity autocorrelation function, presented in
Fig.~\ref{Tmode(omega)}. The central one, localized at $\omega=0$,
is related to the processes of adsorbate site-to-site random
hopping. This peak is determined by the incoherent contribution
\bea\label{Din-t} D_{in}(t)=(a t_{1}/h)^2\mbox{Re}\left[
\gamma_c(t)+\gamma_p(t) \right] \eea to the generalized diffusion
coefficient. In a weak coup\-ling limit, at hight temperatures,
and at large times the end-changing kinetic kernels decay as
\bea\label{PhicApprox}
\gamma_c(\tau)\sim\exp\left[-\eta_c|\ln\omega_0|\left(k_B
T\tau^2+i\tau \right) \right],\eea while the end-preserving ones
decay as \bea\label{PhipApprox1} \gamma_p(\tau)\sim
\Theta(\omega_0^{-1}-t)\left(1/\tau^{2\eta_p k_B
T}-\omega_0^{2\eta_p k_B T}\right).\eea It is seen from
Eqs.~(\ref{PhicApprox})-(\ref{PhipApprox1}) that the relaxation
law of the kinetic kernels changes from the Gaussian behavior for
the sub-Ohmic case with $n=0$ to the truncated long tails for the
super-Ohmic case with $n=2$. It is also easy to show that the
kinetic kernels in the Ohmic regime behave as
\bea\label{PhiOhmicApprox} \gamma_c(\tau)\sim\exp(-\pi\eta_c  T
\tau).\eea

In a weak coupling limit and at the vibrational frequency large
enough there is also an additional maximum of $\mbox{Re}[\tilde
C_{vv}(\omega)]$. This peak is determined by the adsorbate
scattering from the substrate atoms \cite{JCP2} and corresponds to
recrossing processes dealt with the adparticle nonmonotonic
dissipative motion \cite{PRE2009,PRE2011}. Hence, a side peak
arises due to the coherent contribution to the generalized surface
diffusion coefficient (\ref{Dgen1}). The expression for frequency
dependence of the T-mode, associated with the above mentioned
maximum of $\mbox{Re}[\tilde C_{vv}(\omega)]$ can be easily
obtained from Eq.~(\ref{Dgen1}), putting $z=-i\omega+0^+$
\bea\label{TmodeGen} \nonumber &&\tilde
C_{vv}^{coh}(\omega)\equiv\mbox{Re}\left[\tilde
D_{coh}(\omega)\right]\\
&&= \frac{a^2 t_{inter}^2}{4\hbar^2} \frac{2 R(\omega)}{\left[
\omega-\Omega^2 I(\omega)\right ]^2+\Omega^4
R(\omega)^2},\\
\nonumber &&
I(\omega)=\int\limits_0^{\infty}\gamma_c(t)\sin(\omega t)dt,\quad
 R(\omega)=\int\limits_0^{\infty}\gamma_c(t)\cos(\omega
t)dt. \eea It is seen from Eq.~(\ref{TmodeGen}) that the T-mode
localization in the vicinity of the vibrational frequency $\Omega$
is determined by $I(\omega)$, whereas the T-mode width depends on
$R(\omega)$.

Let us study the time dependence of the coherent contribution to
the generalized diffusion coefficient in more detail. Performing
the inverse Laplace transformation, one can obtain the following
expression: \bea\label{inverseL}\nonumber &&
\!\!\!D_{coh}(t)=\!\mbox{Re}\!\left[ \frac{(a t_{inter})^2}{2\pi
i\hbar^2}\lim_{\epsilon\to
0}\!\!\int\limits_{\epsilon-i\infty}^{\epsilon+i\infty}\!\!
dz\exp(z
t)\frac{1}{z+\Omega^2\tilde{\gamma}_{c}(z)}\right] \\
&&\!\!\!=\left(\frac{a t_{inter}}{\hbar}\right)^2\mbox{Re}\left[
\sum_{i=1}^{\infty}\exp(z_i
t)\frac{1}{1+\Omega^2\tilde{\gamma}'_{c}(z_i)}\right].\eea  The
summation in (\ref{inverseL}) in accordance with the residue
theorem runs over all poles $z_i$ of the integrand, which obey the
condition $\mbox{Re} [z_i]<0$. In a general case, the summation is
extended to the infinite number of poles, and the major
contribution comes from terms with maxi\-mal values of
$\mbox{Re}[z_i]$ and weight factors
$[1+\tilde{\gamma}'_c(z_i)]^{-1}$.

The expression for $D_{coh}(t)$ can be even more complicated if
one deals with higher order poles. Indeed, when the temperature
increases, the 2-nd order poles (with rather small weight factors)
appear in the integrand. Even though the high order contributions
to $D_{coh}(t)$ have a form similar to anharmonic corrections to
the velocity autocorrelation function \cite{T-mode-1}, their
physical nature is definitely different. Basically, they are of
non-Markovian origin, while a skewness of the T-mode peak in
Ref.~\cite{T-mode-1} arises due to the anharmonicity of the local
lattice potential.

Just as relaxation times of the incoherent contributions
$D_{in}(t)$ to the generalized diffusion coefficients depend on
the ohmicity of the system (see
Eqs.~(\ref{Din-t})-(\ref{PhiOhmicApprox})), the damping rates
$|\mbox{Re}\,z_i|$ in (\ref{inverseL}) also change over the large
range of values depending on nature of the
``adsorbate--substrate'' interaction. An additional study shows
that at fixed parameters of the model there is a strong
inequality, $\mbox{Max}\{\mbox{Re}[z_i(n=0)]\}\ll
\mbox{Max}\{\mbox{Re}[z_i(n=1)]\}\ll\mbox{Max}\{\mbox{
Re}[z_i(n=2)]\}$, $\mbox{Re}[z_i(n)]<0$, ordering the damping
rates of the coherent term $D_{coh}(t)$ subject to the ohmicity
index $n$.

Turning back to Fig.~\ref{Tmode(omega)}, let us point out that the
T-mode width, which defines the inverse life-time of this
one-particle excitation, can increase for two different reasons.
At fixed temperature the T-peak is broadened when the
``adsorbate--substrate'' coupling increases, and the coherent
motion of the adparticle diminishes. Contrary, one can keep the
coupling fixed and increase the system temperature, ``switching
on'' more and more bath oscillators until their number reaches the
maximum at the Debye temperature $T_D=\hbar\omega_{max}/k_B$. In
such a case not only the T-mode width increases due to higher
dissipation, but also the peak position shifts with temperature.

\begin{figure}[htb]
\centerline{\includegraphics[width=0.25\textwidth]{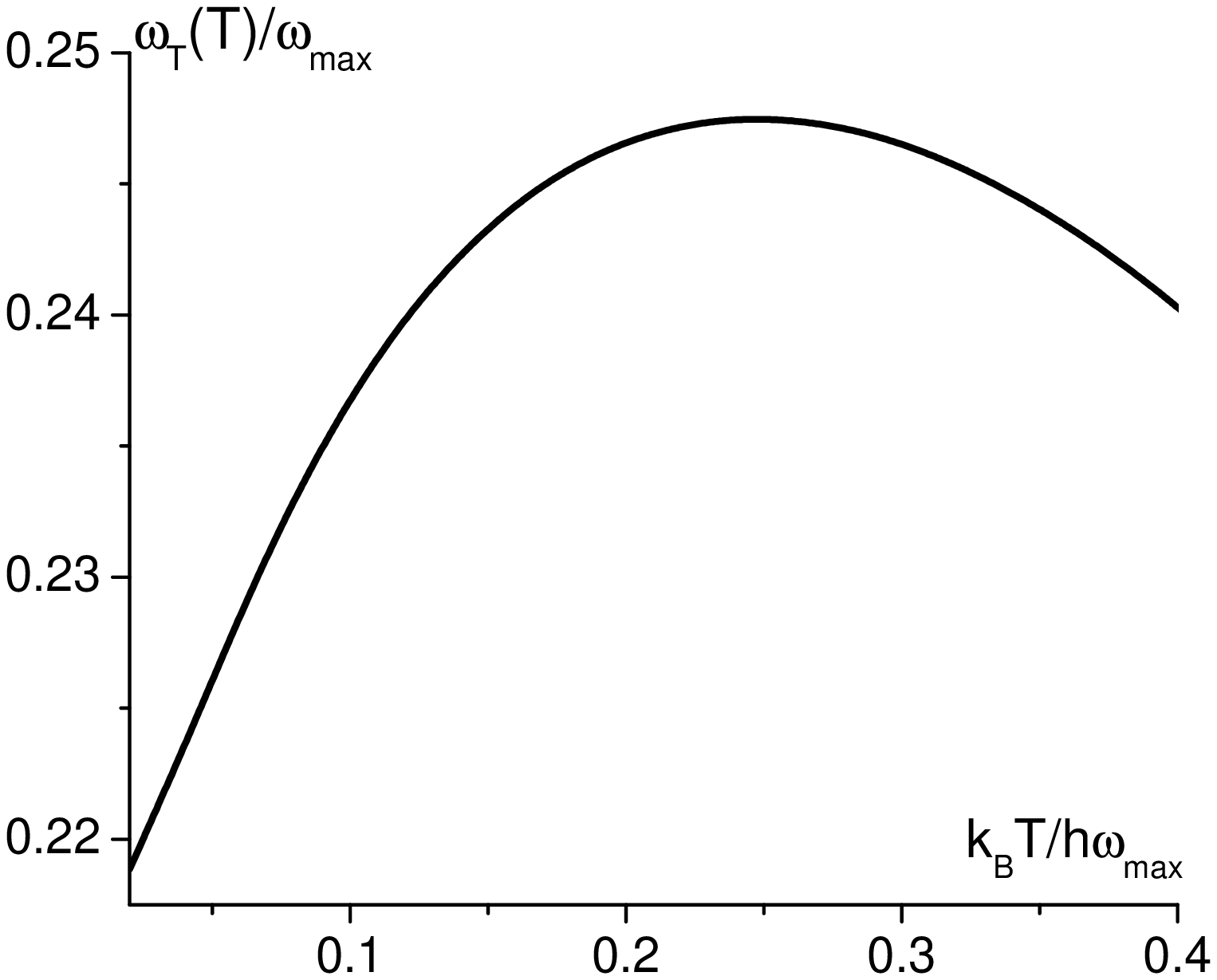}
\includegraphics[width=0.25\textwidth]{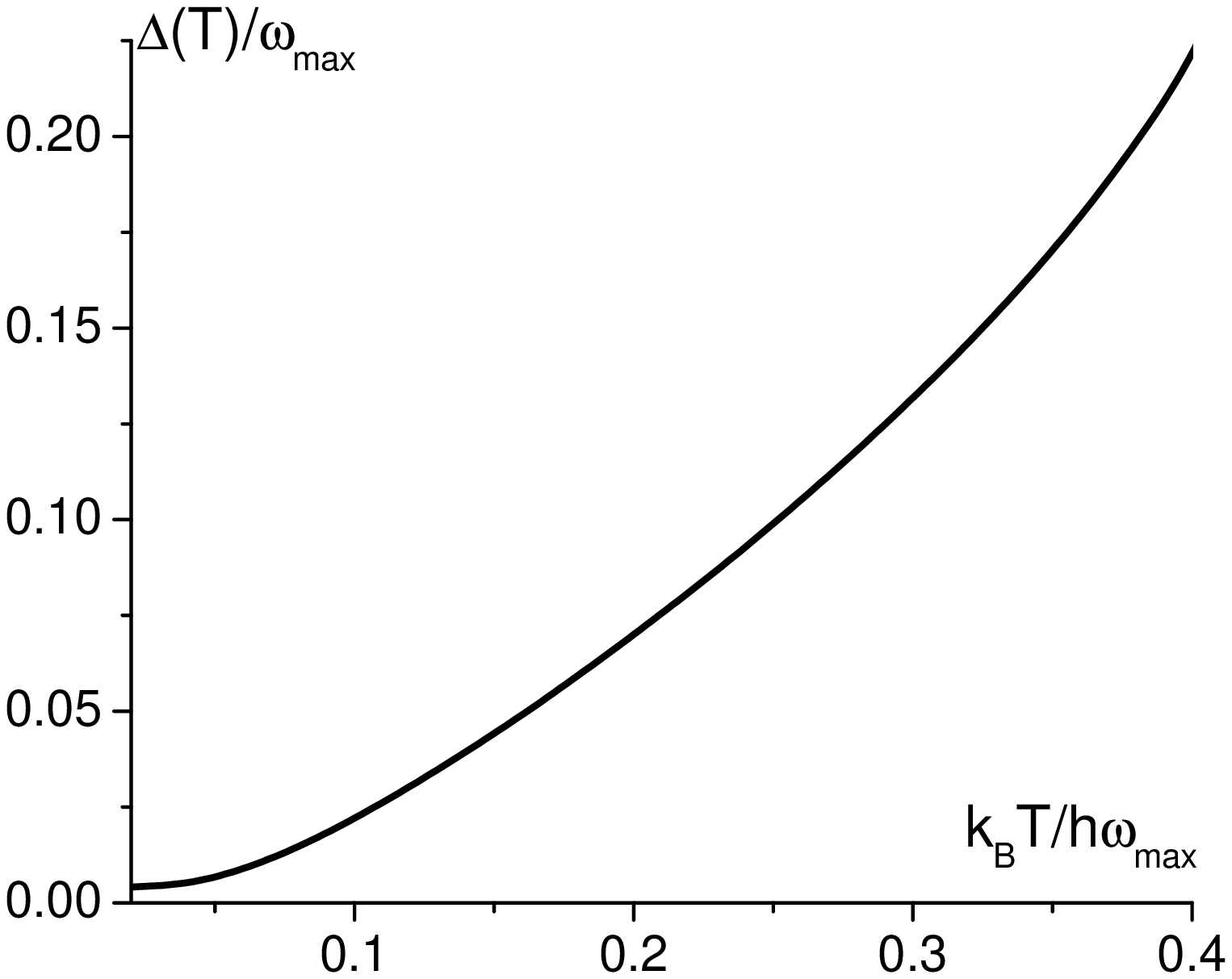}}
\caption{Left panel: temperature dependence of the
T-mode frequency in the sub-Ohmic regime with $n=0.03$,
dimensionless vibrational frequency $\Omega/\omega_{max}=0.2$ and
coupling constant $\eta_c=0.002$. Right panel: temperature
dependence of the T-mode FWHM at the same parameters.}
\label{sub-Ohmic}
\end{figure}

\begin{figure}[htb]
\centerline{\includegraphics[width=0.25\textwidth]{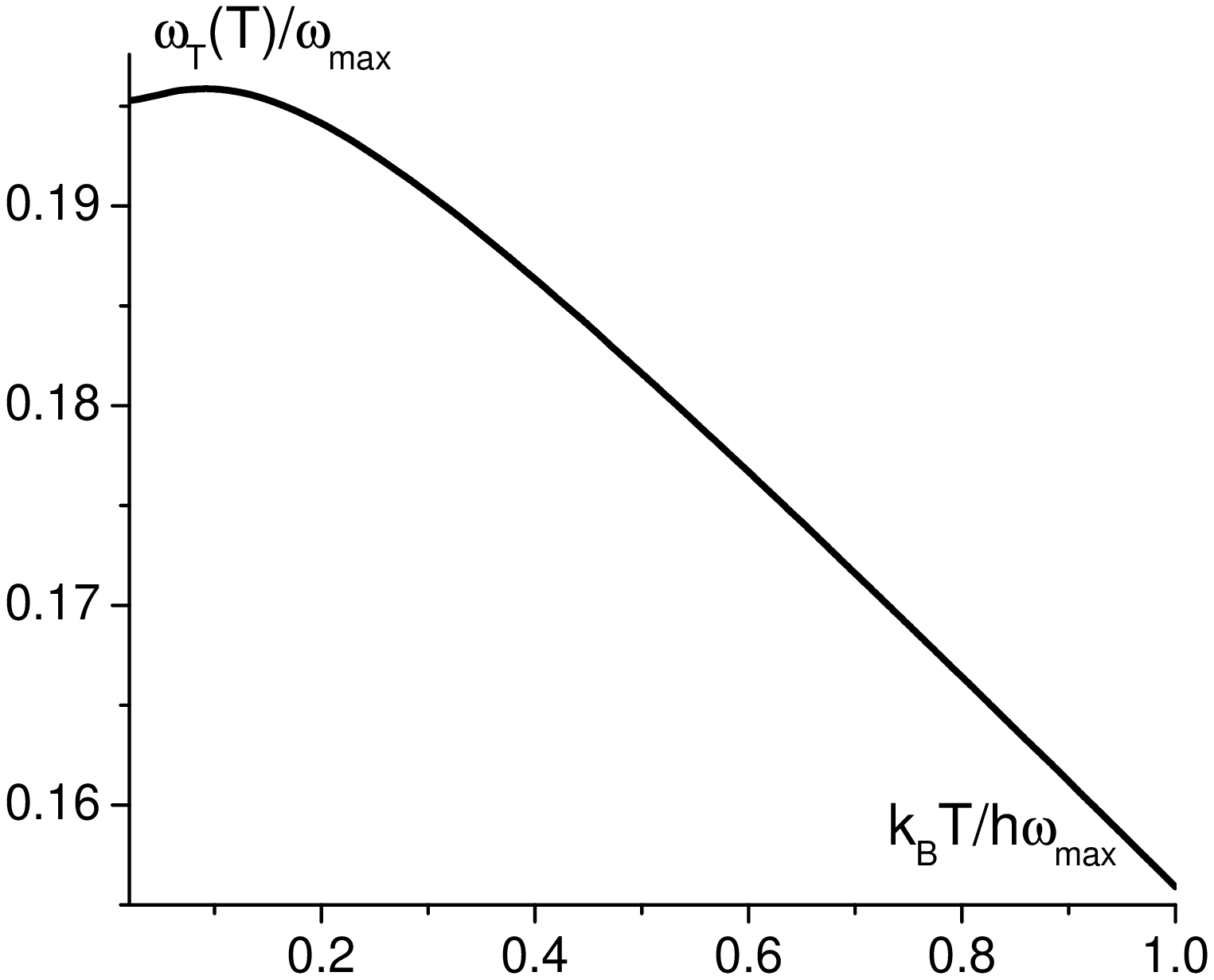}
\includegraphics[width=0.25\textwidth]{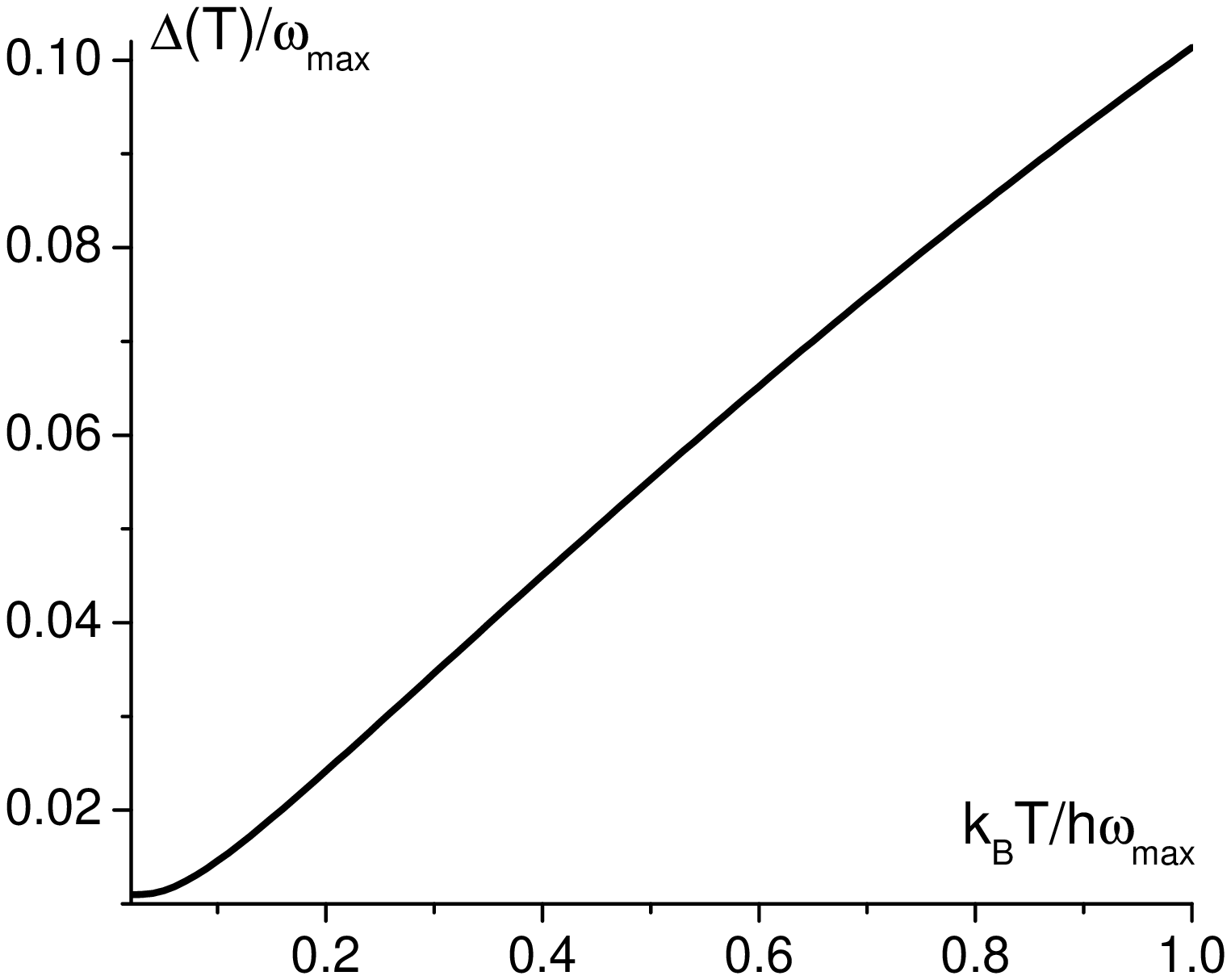}}
\caption{Left panel: temperature dependence of the T-mode
frequency in the super-Ohmic regime with $n=1.97$, dimensionless
vibrational frequency $\Omega/\omega_{max}=0.2$ and coupling
constant $\eta_c=0.166$. Right panel: temperature dependence of
the T-mode FWHM at the same parameters.} \label{super-Ohmic}
\end{figure}

\begin{figure}[htb]
\centerline{\includegraphics[width=0.25\textwidth]{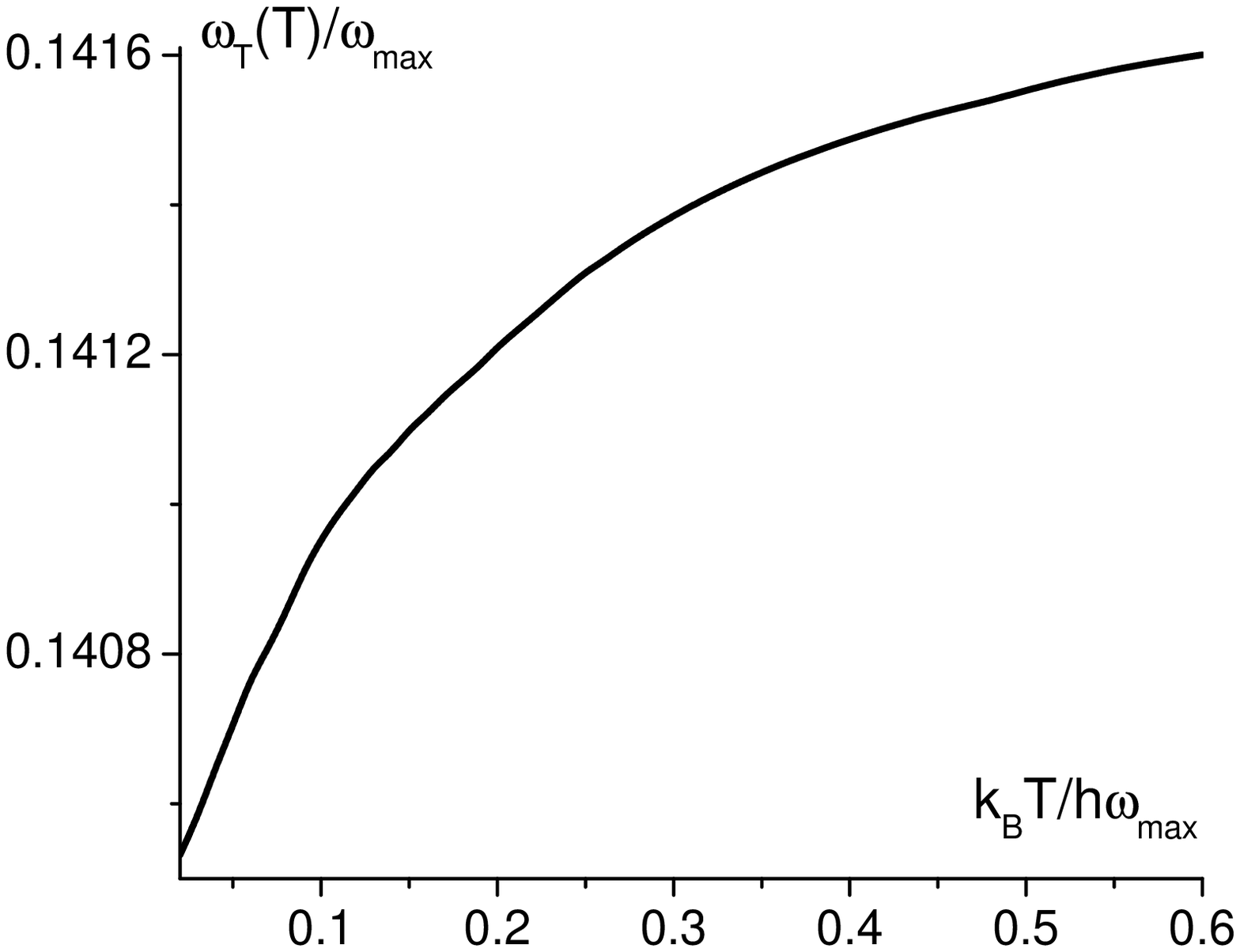}
\includegraphics[width=0.25\textwidth]{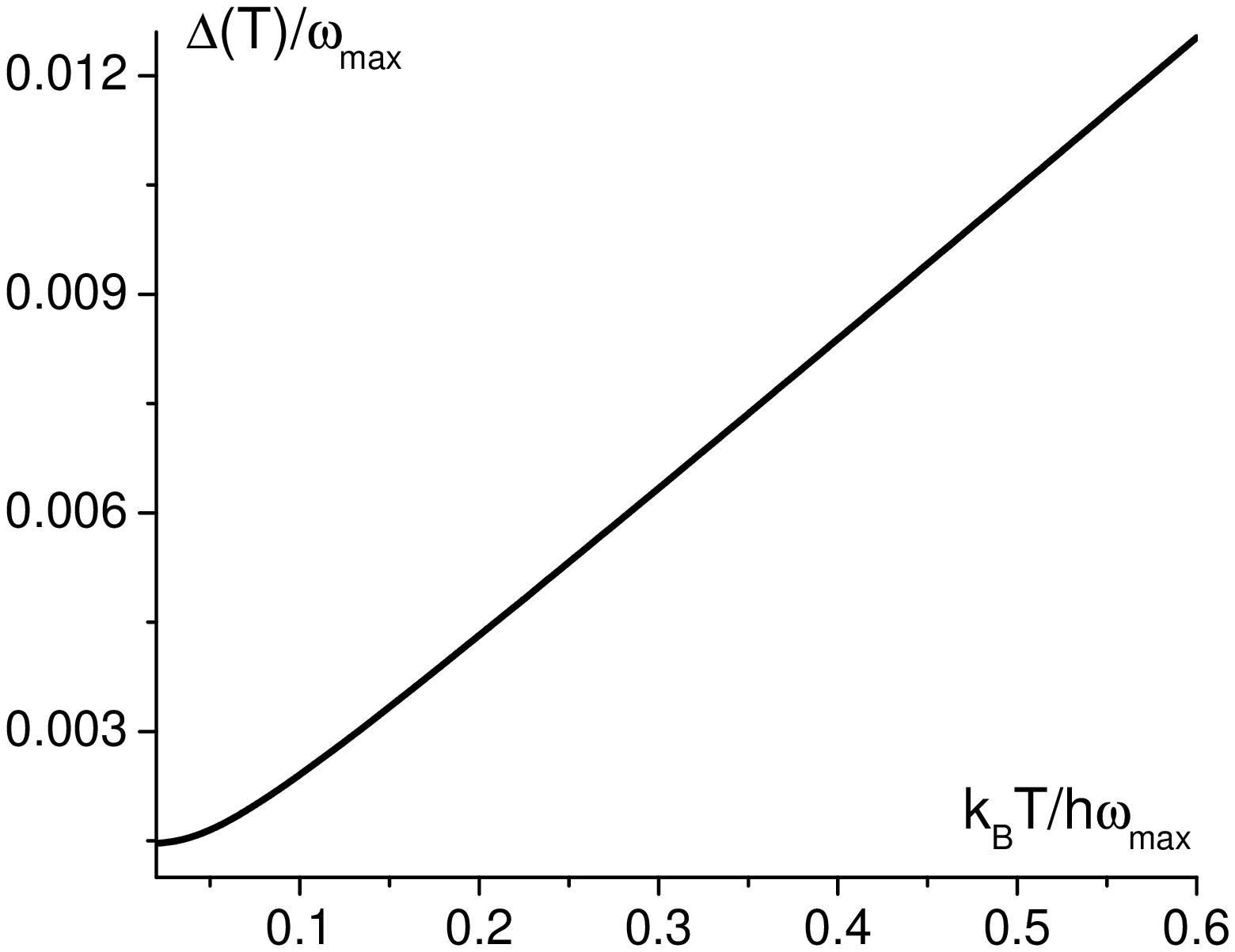}}
\caption{Left panel: temperature dependence of the T-mode
frequency in the Ohmic regime with dimensionless vibrational
frequency $\Omega/\omega_{max}=0.141$ and coupling constant
$\eta_c=0.0066$. Right panel: temperature dependence of the T-mode
FWHM at the same parameters.} \label{Ohmic}
\end{figure}

The temperature dependence of the T-mode frequency and its FWHM
are shown in Figs.~\ref{sub-Ohmic}-\ref{Ohmic}. In the sub- and
super-Ohmic regimes we used the power indexes $n=0.03$ and
$n=1.97$ (which slightly deviate from the corresponding integer
values 0 and 2) to eliminate the lower bound cut-off frequency
$\omega_0$. The others fitting para\-meters (the vibrational
frequency $\Omega$ and the coupling strength $\eta_c$) are chosen
to reproduce the experimental data \cite{T-mode-2}.

It is seen from Fig.~\ref{sub-Ohmic} that in a strong sub-Ohmic
regime the T-mode frequency is {\it blue}-shifted with $T$ up to
temperatures of about $\hbar\Omega/k_B$, where the T-mode
frequency becomes a descending function of temperature.

The T-mode FWHM is found to be a nonlinear function of $T$ at all
temperatures studied. At $k_B T/\hbar\omega_{max}\sim 0.4$ the
T-mode becomes very broad in spite of very small coup\-ling, and
almost disappears from the spectrum function $\mbox{Re}[\tilde
C_{vv}(\omega)]$ (see next Section for analytic estimations).

To explain the T-mode {\it blue}-shift at low-to-moderate
temperatures let us recall that in the sub-Ohmic regime with $n=0$
there are residual correlations of the flicker noise \cite{YK}.
The correlation time $\tau_{cor}$ of the thermal bath random
forces is much larger than the relaxation time $\tau_{v}$ of the
velocity autocorrelation functions (see Eqs.~(\ref{PhicApprox}),
(\ref{inverseL})). Thus, a non-relaxed energy of the lattice is
being delivered to the adparticle increasing its effective
vibrational energy until the inverse process begins at the
temperatures of about $\hbar\Omega/k_B$.

In Fig.~\ref {super-Ohmic} we present the results for the T-mode
tempe\-ra\-tu\-re induced shift and broadening in the strong
super-Ohmic regime. There are ultra-short range time correlations
$\tau_{cor}$ of the thermal bath random forces at super-Ohmic
coupling \cite{Weiss}, and much longer times $\tau_v$ at which the
velocity of the adparticle decays (see Eqs.~(\ref{PhipApprox1}),
(\ref{inverseL})). In such a case the excess energy of the
adsorbate is being transferred to the lattice, yielding a decrease
of the effective vibrational energy of the adparticle with
temperature and, consequently, the {\it red}-shift of the T-mode
frequency. The FWHM of the T-mode accommodates to almost a linear
function of temperature after a short transition regime at low
$T$.

At the Ohmic coupling (see Fig.~\ref{Ohmic}) the T-mode is
slightly {\it blue}-shifted with temperature, and its FWHM is a
linear function of $T$ like in the super-Ohmic case. At the Ohmic
coupling the difference between two typical times $\tau_{cor}$ and
$\tau_v$ is smaller than in the other regimes. As a result, the
T-mode shift is also smaller than in the sub- or super-Ohmic
regimes (at the same values of the parameters of the model), which
will be shown in the next Section. Such a behavior is consistent
with the experimental data of Ref.~\cite{Gadzuk} where the e-h
pair contribution (with a weak Ohmic coupling \cite{MorozovBook})
was found to be almost independent of temperature.

Summing up this Section, we would like to discuss the difference
between our approach and the one used in Ref.~\cite{T-mode-2} when
studying the temperature behavior of the adsorbate T-mode. In
Ref.~\cite{T-mode-2} two fitting parameters were used, which
characterize the behavior of the spectral weight function
$J_x(\omega)$ times a thermal population factor in the vicinity of
the T-mode frequency. These parameters were found to depend very
weakly on the surface temperature for the three systems studied.
Moreover, the dynamic structure factor itself has been calculated
in the Wigner-Weisskopff approximation \cite{Micha19}. That
introduced an additional inaccuracy in the description of the
system dynamics at intermediate times, when the memory effects
have to be taken into account rigorously \cite{PRE2009,PRE2011}.

Even though in Ref.~\cite{T-mode-2} the quantum model for the
adsorbate motion has been proposed, further calculations were
carried out in the impulsive collision approximation without
reference to the driving Hamiltonian term. The influence of
non-selective measurements (e.g. collisions) on the evolution of
quantum system was considered more precisely in \cite{Zeno}, where
the short- and long-time dynamics of the adsorbate has been
studied in the context of quantum Zeno and anti-Zeno effects,
however, without taking into account the tunnelling-mediated
surface diffusion and memory effects.

In our paper we consider a more general case when the ``driving
force'' is dealt with slow underbarrier site-to-site hopping of
the adsorbate and should be treated on equal footing with other
interaction mechanisms. The interaction term (\ref{H_int}) of
Hamiltonian is not of a bilinear form in creation/annihilation
operators. It takes into account coupling both to density and
oscillation modes of the adsorbate that corresponds to more
realistic systems \cite{Ferrando,JCP1}. Neither a power series
expansion for the spectral weight functions at vibration frequency
$\Omega$ nor any approximation for the kinetic kernels (like the
Wigner-Weisskopff one) are made in our paper. As a result, we
manage to describe the T-mode behavior quite accurately using only
one fitting parameter - the dimensionless coupling constant. It is
shown in the next Section that its value can be extracted from the
residual \cite{T-mode-2} shift and broadening of the T-mode at
zero temperature.

\section{An analytical evaluation of the  T-mode temperature dependence}

\setcounter{equation}{0}
Analytical results can be obtained in the
zero temperature and high temperature limits. In the first case, a
thermal population factor from Eq.~(\ref{phi}) is eliminated. On
the other hand, at high temperatures the thermal population factor
can be approximated as $\coth(\hbar\omega/2 k_B T)\approx 2 k_B
T/\hbar\omega$. Then using the long time asymptotics
(\ref{PhicApprox})-(\ref{PhiOhmicApprox}) for the kinetic kernels
instead of the exact expressions (\ref{Gamx}) we can obtain
analytical expressions for the T-mode frequency and FWHM.

\subsection{Zero temperature limit}
In the Ohmic regime and in the weak coupling limit $\eta_c\ll 1$
the following expression for the T-mode frequency is valid
\bea\label{zeroW} \omega_T(0)=\sqrt{\Omega^2- \Delta(0)^2/4}, \eea
that coincides with the expression for the frequency of the damped
harmonic oscillator. The expression (\ref{zeroW}) for the T-mode
localization frequency defines a shift that can be referred to as
the residual one, insofar it is present even at 0 K. The
corresponding residual (or intrinsic) FWHM in the sub- and Ohmic
regimes ($n\le 1$) is scaled as \bea\label{zeroD}
\Delta(0)\sim\eta_c\Omega^n. \eea The damping coefficient (\ref
{zeroD}) is not zero in the $T\to 0$ limit, because the excited
adsorbate can transfer its energy to the lattice, induce
excitations from the zero-point motions of the lattice modes, or
create e-h excitations in the electron distribution.

Let us note that the scaling law (\ref {zeroD}) for the T-mode
FWHM agrees completely with the results of Ref.~\cite{T-mode-2}
obtained in the zero temperature limit. Besides, a practical
significance of such results becomes quite clear: inspect\-ing the
T-mode localization, one can draw a conclusion about the
vibrational frequency $\Omega$ whereas FWHM includes the
information about the ``adsorbate--substrate'' coup\-ling strength
$\eta_c$.

Our model does not admit a straight zero temperature limit in the
super-Ohmic regime because there is no polaron band narrowing in
such a case \cite{JCP1}, and a new derivation of the coherent term
of the generalized diffusion coefficient becomes indispensable.
Nevertheless, the obtained results are quite reliable to be used
for the analysis of the low temperature T-mode features from both
theoretical and experimental viewpoints.

\subsection{High temperature limit}

\subsubsection{Sub-Ohmic regime with $n=0$}
Taking into account the Gaussian form (\ref{PhicApprox}) of the
long time asymptotics for the kinetic kernels and
Eq.~(\ref{TmodeGen}) for the frequency dependence of the T-mode it
is straightforward to obtain the expansion for the T-mode
frequency \bea\label{W_T_0} \omega_T(T)=\Omega+\frac{\eta_c T |\ln
\omega_0|}{\Omega}\left\{1-\frac{5}{2}\frac{\eta_c T |\ln
\omega_0|}{\Omega^2} \right\} \eea as a series in the coupling
strength $\eta_c$. The corresponding FWHM $\Delta(T)$ can be
written down as \bea\label{D_T_0} \Delta(T)=\frac{\Omega^2
\exp(-\Omega^2/4 \eta_c T |\ln\omega_0|)}{2\sqrt{\eta_c T
|\ln\omega_0|}}; \eea it does not admit the $\eta_c$ series
expansion. The T-mode FWHM is found to be a nonlinear function of
temperature, which agrees completely with the results presented in
Fig.~\ref{sub-Ohmic}.

Analyzing the T-mode temperature dependence (\ref{W_T_0}) it is
possible to find the temperature \bea\label{Tstar}
T^*=\frac{\Omega^2}{5\eta_c|\ln\omega_0|}, \eea at which a maximal
shift of the T-mode \bea\label{maxShift0}
\Delta\omega_T(T^*)\equiv\omega_T(T^*)-\Omega=\frac{1}{10} \Omega
\eea takes place. The shift (\ref{maxShift0}) is quite large, and
the T-mode FWHM at this temperature \bea\label{maxD0}
\Delta(T^*)=\sqrt\frac{5}{4}\exp\left(-\frac{5}{4}\right)\Omega\approx\frac{1}{3}\Omega
\eea approaches the vibrational frequency, leading to a rapid
decay of the excitation as it is clearly seen in
Fig.~\ref{sub-Ohmic}.

\subsubsection{Ohmic regime}
Analysis of the T-mode temperature dependence at the Ohmic
coupling shows that at reasonable values of the vibrational
frequency $\Omega$ the T-mode shift
\bea\label{W_T_1}\Delta\omega_T(T)=-\frac{(\pi\eta_c
T)^2}{2\Omega} \eea is smaller than in sub- or super-Ohmic
systems, since the expansion in (\ref{W_T_1}) starts from a
quadratic term in the coupling strength. The FWHM is governed by
the expression \be\label{D_T_1} \Delta(T)\sim\eta_c T\Omega^{n-1},
\ee which is also valid for the super-Ohmic systems with $1<n\le
2$.

It should be stressed that the temperature dependence of the
T-mode frequency (\ref{W_T_1}) differs from the results presented
in Fig.~\ref{Ohmic}. The T-mode is {\it red}-shifted in accordance
with Eq.~(\ref{W_T_1}), whereas the {\it blue}-shift is observed
in Fig.~\ref{Ohmic}. An explanation of this contradiction is quite
simple: we used the exponential form (\ref{PhiOhmicApprox}) of the
kinetic kernel when calculating (\ref{W_T_1}), while at the
numerical evaluation we used the exact expression (\ref{Gamx}),
which up to 5-10\% differs from asymptotic value
(\ref{PhiOhmicApprox}) at short times. This deviation can be
considered negligible when calculating the temperature dependence
of the diffusion coefficients \cite{JCP2,PRE2011}. But it is not
true when one evaluates the temperature dependence of one-particle
excitation of the adsorbate, that once more shows the exceptional
sensitivity of the T-mode behavior to any kind of approximations.

\subsubsection{Super-Ohmic regime with $n=2$}

At the super-Ohmic regime with $n=2$, weak coup\-ling limit
$\eta_c\ll 1$, and moderate vibrational frequencies
$\Omega<\omega_{max}$, the temperature dependence of the T-mode
localization can be presented as \bea\label{W_T_2}
\omega_T(T)=\Omega+\eta_c T\Omega\ln(2\Omega). \eea

It is seen from the last expression that the T-mode is {\it
red}-shifted with temperature up to the vibrational frequencies
$\Omega/\omega_{max}=1/2$, being at the same time a strongly
nonlinear function of $\Omega$. Exact numerical evaluations show
that the $\omega_T(T)$-curve at first reaches a plateau as the
vibrational frequency increases and then becomes a decreasing
function of temperature at $\Omega\sim \omega_{max}/2$.

On the other hand, the T-mode FWHM is very small even at the Debye
temperature \bea\label{D_T_2} \Delta(k_B
T/\hbar\omega_{max}=1)\sim\eta_c\Omega\ll\Omega. \eea Thus, the
T-mode is not broadened too much and should be well defined. It is
also to be pointed out that in the super-Ohmic regime the T-mode
shift is much smaller than that in the sub-Ohmic case, since
$\Omega|\ln\Omega|\ll|\ln\omega_0|/\Omega$.

Summarizing, the analytic investigation of the T-mode behavior
confirms completely the main tendencies observed in
Figs.~\ref{sub-Ohmic}-\ref{Ohmic}. Namely: i) a linear increase of
the T-mode FWHM with temperature in the Ohmic and super-Ohmic
regimes and a non-linear increase of $\Delta(T)$ in the sub-Ohmic
regime; ii) the corresponding {\it red}- or {\it blue}-shifts of
the T-mode location; iii) a considerable T-mode broaden\-ing in
the sub-Ohmic regime even at weak ``adsorbate--substrate''
coupling and a small broadening at $1\le n\le 2$.

Since there is a one-to-one correspondence between the system
ohmicity and the kind of the coupling, one can deduce about a
nature of the ``adsorbate--substrate'' interaction from the
temperature induced shift of the T-mode. When several kinds of the
``adsorbate--substrate'' interaction act simultaneously, the data
obtained could be more complicated than those reported in this
Section. However, such a situation is rather infrequent since
usually the typical relaxation times of different interactions are
separated by several orders of magnitude \cite{Ferrando,T-mode-2},
and it is possible to talk about a single effective interaction.

\section{T-mode and multiple adsorbate hopping}

\setcounter{equation}{0}

Investigation of the adparticle diffusion on the basis of the
Klein-Kramers equation \cite{PhysA195,SurScience311,Kramers}
allows us to look at the problem of T-mode from another viewpoint,
considering this one-particle excitation as a precursor of
multiple or long jumps of the adsorbate. Such a jumping regime
should not be confused with the situation, when at very low
barriers the particle performs a quasi-continuous hopping at
several sites \cite{PhysA195,SurScience311}, and a diffusive stage
of evolution is not reached. One  rather has to talk about the
situation when the velocity of the particle, which has localized
at a certain adsorption site $s$, is not thermalized yet, and the
adsorbate jumps to the next nearest neighboring site $s'$ very
soon. Thus, we will use the term ``multiple'' rather than ``long''
hopping of the adsorbate.

In Ref.~\cite{SurScience311} two condi\-tions for the multiple
jumps of the adsorbate at the thermal activated diffusion have
been established, formulated as strong inequalities
\bea\label{condC} \tau_{th}\ll\tau_v,\qquad\tau_{osc}\ll\tau_v
\eea for three typical times of the system dynamics
\bea\label{3timesC} \tau_{osc}=a\sqrt\frac{m}{U},\quad \tau_v\sim
\frac{1}{\eta_c}, \quad \tau_{th}=a\sqrt{\frac{m}{k_B T}}, \eea
where $\tau_{osc}$ means the period of oscillation of the
adparticle with mass $m$ at the bottom of the potential well of
the depth $U$; $\tau_v$ is a relaxation time of the velocity
autocorrelation function, and $\tau_{th}$ denotes the time taken
by the particle to cross over a lattice spacing $a$ with a mean
thermal velocity $v_{th}$.

It is seen from the presented above analysis that the T-mode is
defined by the second inequality of (\ref{condC}). If the system
parameters allow the typical times to obey the first inequality
too, then the multiple jumps can be observed in the system.

Let us suppose that conditions (\ref{condC}) are fulfilled. Then
during both the jump to the nearest neighboring site (left
inequality) and oscillation within a certain adsorption site
(right inequality) the particle does not have enough time to be
thermalized, and the multiple hopping scenario of the diffusion
can be realized. In such a case, the Einstein-Smoluchowski
equation is found to be insufficient \cite{PhysA195,SurScience311}
for the description of the adparticle diffusion, and one has to
use the Klein-Kramers equation for the distribution function,
which depends both on the velocity and on the coordinate of the
adsorbate.

In the case of quantum diffusion the multiple hopping conditions
(\ref{condC}) remain valid too, but the values $\tau_{osc}$ and
$\tau_{th}$ from Eq.~(\ref{3timesC}) should be substituted by the
typical times \bea\label{2timesQ}
\tau_{osc}=\frac{2\pi}{\Omega},\qquad
\tau_{tun}=\frac{2\pi\hbar}{t_{inter}},\qquad\eea where
$\tau_{osc}$ denotes the inverse vibrational frequency, and
$\tau_{tun}$ is a time taken by the particle to perform an
underbarrier hopping to the nearest neighboring site.

It should be pointed out that Refs.~\cite{PhysA195,SurScience311}
were directed mainly to the investigation of inelastic peak of the
dynamic structure factor (or, alternatively, the side peak of
velocity autocorrelation function) at various coupling constants
and barrier heights, and no study of the temperature dependence of
the inelastic peak was performed to attribute it to the T-mode
features.

On the other hand, a generalization of the Klein-Kramers equation
to description of the quantum diffusion (especially in the weak
coupling limit, when there is the energy-diffusion-controlled
regime from the viewpoint of transition state theory
\cite{Kramers}) is an intricate problem, which, up to our
knowledge, has not been solved yet. In Ref.~\cite{PRE2011} we
followed an alternative way, having studied a behavior of the
``velocity--velocity'' time correlation functions, defined on the
adjacent sites, with taking into account the memory effects. It
has been shown that at given parameters of the model the multiple
jumps are absent (the relation $\tau_{tun}\sim\tau_v$ was found to
be valid), but they should appear when the tunnelling amplitude
increases or the coupling strength decreases. The most interesting
question -- about the influence of temperature on the multiple
hopping of adsorbate -- can be answered only after an additional
analysis of the T-mode behavior.

\section{Concluding remarks}
In this paper, we carry out numerical and analytical investigation
of the temperature dependence of the T-mode of adsorbate which
diffuses from one adsorption site to another by a tunnelling
mechanism. The T-mode characteristics (its localization frequency
and width) are known to depend on temperature
\cite{T-mode-1,T-mode-2,PRE2011}. Having adopted a concept that
the T-mode appearance is dealt with memory effects of the
adsorbate motion, we propose an explanation of either ({\it red}
or {\it blue}) temperature induced shifts of the above mentioned
one-particle excitation on the basis of comparative analysis of
two typical timescales: the correlation time $\tau_{cor}$ of the
thermal bath random forces and the relaxation time $\tau_v$ of the
velocity autocorrelation functions.

We show that in the strong sub-Ohmic regime with $n\ll 1$ the
relation $\tau_v\ll\tau_{cor}$ is valid, leading to a {\it
blue}-shift of the T-mode with temperature, since the excess
energy of the lattice is being delivered to the adparticle,
thereby increasing its effective vibrational energy. On the
contrary, in the strong super-Ohmic regime with $n\approx 2$ the
inequality $\tau_v\gg\tau_{cor}$ becomes valid. At the timescales
of about $\tau_{cor}$ the adparticle does not have enough time to
be thermalized and starts to transfer its energy to the lattice,
yielding the observed {\it red}-shift of the T-mode.

We have also analyzed in detail the temperature behavior of the
T-mode width, which changes from a strongly nonlinear function of
$T$ (in the case of strong sub-Ohmic regime with $n\ll 1$) to the
linear dependence in temperature (for Ohmic and super-Ohmic
systems).

The ohmicity type is closely related to the nature of the
``adsorbate--substrate'' interaction: the case with $n=0$
corresponds to the flicker noise, the Ohmic regime is being
realized in the systems with ``electronic'' friction or at the
coupling with nonlinear phonons, while the super-Ohmic case with
$n=2$ corresponds to the linear Debye phonon coupling. Therefore,
one can draw a conclusion about the interaction mechanism from
analysis of the temperature induced shift of the T-mode. On the
other hand, extrapolating the obtained data to the zero
temperature limit, it is possible to evaluate the micro\-scopic
parameters (vibrational frequency $\Omega$ and coupling strength
$\eta_c$), which govern behavior of the ``adsorbate--substrate''
system.

We consider a slow underbarrier site-to-site hopping of the
adparticle to be a ``driving force'' of the surface motion of the
adsorbate. Unlike the impulsive collision approximation
\cite{T-mode-2}, a ``driving term'' (\ref{H_A}) of the Hamiltonian
is treated on equal footing with other interaction mechanisms and
defines a time evolution of the system. Since there is a lack of
data about the T-mode beha\-vior for the case, when the diffusion
is mediated by tunnelling, and the non-Markovian effects play an
important role \cite{Zeno} (or, what is almost the same, when
different regimes of the ohmicity manifest themselves), we believe
that our investigations have a good perspective from a viewpoint
of the study of transition regimes of the adsorbate.

It has to be emphasized that a similar approach should be quite
promising in the case of thermally activated diffusion as well. It
was shown in Ref.~\cite{Pottier} that in non-Ohmic systems with
$0<n<2$, $n\ne 1$, the thermal bath random forces were correlated
as $\langle F_r(t) F_r(0)\rangle\sim\frac{\eta_c k_B T}{t^n}$,
whereas the long time asymptotics of the velocity autocorrelation
functions was scaled as $\langle v(t) v(0)\rangle\sim\frac{\eta_c
k_B T}{m} \frac{1}{t^{2-n}}$. Thus, the relations
$\tau_v\ll\tau_{cor}$ or $\tau_v\gg\tau_{cor}$ remain valid also
at thermally activated diffusion, allowing one to perform a
similar comparative analysis of the typi\-cal timescales and to
study the temperature behavior of the T-mode.

Finally, we relate the T-mode to the onset of the multihops of the
adsorbate. If an eventual temperature induced shift of the T-mode
toward lower frequencies and its broadening are not too large
(quasi- and inelastic peaks of the dynamic structure factor are
well resolved), one can state that there is a portion of multihops
in the system. The computer simulations within the Monte Carlo
wave function formalism \cite{36Ferrando} and direct evaluation of
the ``flux--flux'' time correlation functions \cite{JJ-TCF,Japan}
support this suggestion. Thus, investigation of the T-mode
behavior along with a study of the ``velocity--velocity'' cross
correlation functions can shed more light upon the nature of such
an interesting phenomenon of the adsorbate motion as a diffusion
by the multi-hopping scenario.

\section*{Acknowledgement}
This work was partially supported by the Project ``Models of the
quantum-statistical description of cata\-ly\-tic processes at the
metallic surfaces''(Lviv Polytechnic National University), No.
0110U001091. The author also thanks Dr. A.Moina for the useful
discussions and the assistance in performing of numerical
evaluations.

\end{document}